\documentclass[letter,onecolumn,10pt]{IEEEtran}

\usepackage{amssymb}
\usepackage{amsmath}
\usepackage{multirow}
\usepackage{epsfig}
\usepackage{rotating}
\usepackage{subfigure}
\usepackage{algorithmic}
\usepackage{algorithm}
\usepackage{psfrag}
\usepackage{graphics}
\usepackage{subfigure}
\usepackage{url}

\newcommand{\subE}{{\mathrm{E}}}
\newcommand{\subBE}{{\mathrm{BE}}}
\newcommand{\subBO}{{\mathrm{BO}}}
\newcommand{\subEO}{{\mathrm{EO}}}

\newfont{\greek}{psyr}

\newtheorem{theorem}{Theorem}[section]
\newtheorem{corollary}[theorem]{Corollary}
\newtheorem{lemma}[theorem]{Lemma}
\newtheorem{proposition}[theorem]{Proposition}
\newtheorem{definition}[theorem]{Definition}

\newtheorem{example}{Example}[section]

\include{PrintXfigPS}
\def\ve#1{{\mathchoice{\mbox{\boldmath$\displaystyle #1$}}%
              {\mbox{\boldmath$\textstyle #1$}}%
              {\mbox{\boldmath$\scriptstyle #1$}}%
              {\mbox{\boldmath$\scriptscriptstyle #1$}}}}

\newcommand{\PG}{{\mathrm{PG}}}
\newcommand{\GField}{{\mathrm{GF}}}
\newcommand{\ld}{{\mathrm{ld}}}

\begin{document}
\title{The Trapping Redundancy of Linear Block Codes 
\thanks{Part of this work was presented at the International
Conference on Wireless Networks, Communications, and Mobile
Computing (WirelessComm) 2005, Maui, Hawaii, at the IEEE Global
Telecommunications Conference (GlobeCom) 2006, San Francisco,
California, and at the Conference on Information Sciences and
Systems (CISS) 2006, Princeton, New Jersey, USA. This work was
supported in part by NSF Grant CCF-0514921 awarded to Olgica
Milenkovic, by a research fellowship from the Institute for
Information Transmission, University of Erlangen-Nuremberg,
Erlangen, Germany, awarded to Stefan Laendner, and by a German
Academic Exchange Service (DAAD) fellowship awarded to Thorsten
Hehn.} 
}

\author{Stefan L{\"a}ndner\ddag, Thorsten Hehn\ddag, Olgica Milenkovic\dag, and
Johannes B. Huber\ddag,\\
\ddag University of Erlangen-Nuremberg, Erlangen, Germany\\
\dag University of Illinois at Urbana-Champaign, Urbana, IL, USA}

\date{}
\maketitle
\thispagestyle{empty}

\begin{abstract}
\label{sec:abstract} We generalize the notion of the stopping
redundancy in order to study the smallest size of a trapping set
in Tanner graphs of linear block codes. In this context, we
introduce the notion of the trapping redundancy of a code, which
quantifies the relationship between the number of redundant rows
in any parity-check matrix of a given code and the size of its
smallest trapping set. Trapping sets with certain parameter sizes
are known to cause error-floors in the performance curves of
iterative belief propagation decoders, and it is therefore
important to identify decoding matrices that avoid such sets.
Bounds on the trapping redundancy are obtained using probabilistic
and constructive methods, and the analysis covers both general and
elementary trapping sets. Numerical values for these bounds are
computed for the $[2640,1320]$ Margulis code and the class of
projective geometry codes, and compared with some new
code-specific trapping set size estimates.
\end{abstract}
\textbf{Index Terms} Belief Propagation, LDPC Codes, Margulis
Codes, Projective Geometry Codes, Trapping Redundancy, Trapping Sets.

\section{Introduction}

The performance of linear error-correcting codes (and low-density
parity-check (LDPC) codes in particular) under iterative decoding
depends on the choice of the parity-check matrix of the code. More
precisely, the error rate of a code is influenced by a class of
combinatorial entities determined by the choice of the
parity-check matrix, such as stopping~\cite{dietal02} and trapping
sets~\cite{mackayetal03,richardson03}. Stopping and trapping sets
are defined in terms of constraints on the weights of rows in the
parity-check matrix induced by subsets of its columns. Certain
such restrictions on the weight distributions of the rows can only
be satisfied if the parity-check matrix of the code has a
sufficiently large number of judiciously chosen rows. Thus, recent
work focused on introducing redundant rows into parity-check
matrices of a code in order to ensure that the size of their
smallest stopping sets are sufficiently large or equal to the
minimum distance of the
code~\cite{hanetal07,koetteretal06,schwartzetal06,weberetal05}.
Since adding redundant rows to the parity-check matrix increases
the decoding complexity of the code, it is important to understand
the inherent trade-off between the size of the smallest stopping
set and the number of rows in a parity-check matrix. Several
ideas for addressing these issues that exploit properties of
orthogonal arrays and covering arrays~\cite{hedayatetal99} were
described in~\cite{schwartzetal06}, \cite{hanetal07}, and
\cite{milenkovicetal06}.

The contributions of this work are three-fold. First, we
generalize the notion of the stopping redundancy for the case of
trapping sets, and term this combinatorial number the
\emph{trapping redundancy}. Second, we describe simple
probabilistic and deterministic methods for upper-bounding the
trapping redundancy of binary linear block codes. Third, we
present new analytical techniques for estimating the sizes of
small trapping sets in the family of projective geometry (PG)
codes and the Margulis $[2640,1320]$ code, and compare these
estimates with the upper bounds.

The paper is organized as follows. Section~\ref{sec:definitions}
provides relevant definitions and introduces the terminology used
throughout the paper. Section~\ref{sec:analytical_study} contains
the main results -- probabilistic and constructive upper bounds on
the trapping redundancy of codes.
Section~\ref{sec:analytical_comparisons} describes the
relationship between trapping sets and arcs in PG codes. In the
same section, numerical results for the trapping redundancy of the
Margulis $[2640,1320]$ and the family of PG codes are compared
with results concerning arcs and elementary trapping sets in the
family of Margulis codes. Concluding remarks are given in
Section~\ref{sec:conclusions}.

\section{Definitions, Background, and Terminology}
\label{sec:definitions}

We start by introducing trapping sets and elementary trapping
sets, and then proceed to define the notions of the restriction of
(redundant) parity-check matrices. Based on the notion of
the restriction, we state the central definition of the paper,
pertaining to the trapping redundancy of a linear block code.

\subsection{Near-Codewords and Trapping Sets}
\label{sec:trapping_sets}

Decoding of LDPC codes is usually performed in an iterative
manner, using the suboptimal belief-propagation (BP) algorithm.
This decoding approach, also known as message passing, can be seen
as the process of exchanging reliability messages along the edges
of a bipartite Tanner graph. The incidence matrix of the Tanner
graph corresponds to the parity-check matrix used for decoding, so
that the code variables index the vertices on the left hand side
of the graph, while the parity-check equations index the right
hand side vertices of the graph. For a more comprehensive
treatment of LDPC codes and iterative decoding, the interested
reader is referred
to~\cite{mackay99,richardsonetal01a,richardsonetal01}.

The error-floor phenomenon of iterative decoders was first
described by MacKay and Postol in~\cite{mackayetal03}, who
observed that the bit error rate curve of the $[2640,1320]$
Margulis code exhibits a sudden change of slope at signal-to-noise
ratios approximately equal to $2.4$ dB. This change of slope was
attributed to the existence of \emph{near-codewords} in the Tanner
graph of the Margulis code with a parity-check matrix $\ve{H}$
described in \cite{rosenthaletal01}. Near codewords are error
vectors $\ve{y}$ of small weight, with syndromes ${\ve{s_y}}
\equiv \ve{H} {\ve{y}}$ that also have small weight. In his
seminal paper~\cite{richardson03}, Richardson analyzed the effect
of near-codewords on the performance of various classes of
decoders and for a group of channels. He also introduced the
notion of \emph{trapping sets} to describe configurations of
variable nodes in Tanner graphs of codes that cause failures of
specific decoding schemes. There exist many different groups of
trapping sets. For example, trapping sets of maximum likelihood
decoders are sets of variables containing the supports of
\emph{codewords} of the code; trapping sets of iterative decoders
used for messages transmitted over the binary erasure channel are
stopping sets~\cite{dietal02}. For the additive white Gaussian
noise (AWGN) channel and 
 BP decoding, no
simple characterization of trapping sets is known. Nevertheless,
extensive computer simulations revealed that a large number of
trapping sets for this channel/decoder combination can be
described in a simple and precise setting. Henceforth, we use the
notion ``trapping set'' to refer exclusively to sets of the form
described below. To define trapping sets, we first introduce the
notion of the restriction of a matrix.

\begin{definition}
For a given $m \times n$ matrix $\ve{H}=(H_{i,j})$ with ${1 \leq i
\leq m}$, $1 \leq j \leq n$, the restriction of a set of $t$
columns indexed by $j_1,j_2,\ldots,j_t$ is defined as an $m \times
t$ sub-matrix of $\ve{H}$ consisting of the elements $H_{i,j}$,\;
${1 \leq i \leq m}$, \; $j=j_1,j_2,\ldots,j_t$.
\end{definition}

For a given linear $[n,k,d]$ code ${\mathcal{C}}$, with
parity-check matrix $\ve{H}$ and corresponding Tanner graph
${\mathcal{G}}(\ve{H})$, trapping sets are defined as follows.

\begin{definition} An $(a,b)$ trapping set $\mathcal{T}(a,b)$ is a
collection of $a$ variable nodes for which the subgraph in
$\mathcal{G}(\ve{H})$ induced by $\mathcal{T}(a,b)$ and its
neighbors contains $b > 0$ odd-degree check nodes\footnote{The
case $b=0$ corresponds to codewords. Henceforth, we consider 
the case $b>0$ only.}. Equivalently, an $(a,b)$ trapping set
$\mathcal{T}(a,b)$ of $\ve{H}$ is a set of $a$ columns with a
restriction that contains $b$ odd-weight rows.
\end{definition}

The class of trapping sets that exhibits the strongest influence
on the performance of iterative decoders is the class of
\emph{elementary} trapping sets.
\begin{definition}
An elementary $(a,b)$ trapping set is a set
$\mathcal{T}^{(e)}(a,b)$ for which all check nodes in the subgraph
induced by $\mathcal{T}^{(e)}(a,b)$ and its neighbors have either
degree one or two, and there are exactly $b$ degree-one check
nodes. Alternatively, an elementary $(a,b)$ trapping set is a
trapping set for which all non-zero rows in the restriction have
either weight one or two, and exactly $b$ rows have weight one.
\end{definition}

For a fixed value of the parameter $a$ (or $b$), the problem of
finding the trapping set with smallest parameter $b$ (or $a$) in a
given parity-check matrix is NP-hard, and NP-hard to
approximate~\cite{mcgregoretal07}. This makes a general and
complete characterization of the trapping set sizes and trapping
redundancy prohibitively complex. We therefore focus our attention
on deriving upper bounds on the trapping redundancy with set sizes
restricted to $a \leq d-1$ only, where $d$ denotes the minimum
distance of the code, and in particular, elementary trapping
sets~\cite{milenkovicetal07}. This is motivated by the recent
studies that suggest that trapping sets most detrimental to the
code performance are elementary, and that they have small $(a,b)$
parameters, usually such that
$b<a<d$~\cite{richardson03,laendneretal05}.

\subsection{Redundant Parity-Check Matrices}

For every linear $[n,k,d]$ code $\mathcal{C}$, there exist many
choices for parity-check matrices, although for iterative decoding
not all of them may be adequate. This is due to the fact that some
parity-check matrices have irregular row- and column-weights and
that they contain a large number of stopping and trapping
sets~\cite{schwartzetal06}. In order to mitigate this problem, one
may resort to the use of \emph{redundant} parity-check matrices,
i.e., matrices that contain more than $n-k$ rows, although they
have row-rank equal to $n-k$. Examples of the use of redundant
parity-check matrices for signaling over the binary erasure
channel can be found
in~\cite{schwartzetal06,hollmannetal07,hehnetal06}.

Henceforth, we use the phrase \emph{redundant parity-check matrix}
to refer to a parity-check matrix with row-rank $n-k$ that has
more than $n-k$ rows. A redundant parity-check matrix contains
rows that are linear combination of other rows that represent a
basis of the dual code $\mathcal{C}^{\perp}$. For a fixed basis,
rows of this form are referred to as redundant rows. On the other
hand, a parity-check matrix of full row-rank and dimension $(n-k)
\times n$ is simply termed a \emph{parity-check matrix}. Redundant
parity-check matrices are used to impose specific constraints on
the structure of their corresponding Tanner graphs. As was shown
in~\cite{laendneretal06}, even one judiciously chosen redundant
row can be used to lower the error floor of the Margulis code.
This is achieved in terms of rendering the structure of a selected
small trapping set so as to increase the number of its
corresponding unsatisfied parity-check equations.
From an application point of view, it is of interest to identify
one or a few redundant rows that can be added to the parity-check
matrix in order to eliminate a trapping set causing a special
instant of decoding failure. Note that one or a few redundant rows
of low weight do not significantly alter the performance of a code
in the waterfall region, although they may have a significant
bearing on its performance in the error-floor region.
In what
follows, we consider the more general
 theoretical
problem of determining the smallest number of redundant rows
needed to \emph{simultaneously} eliminate the negative effect of
classes of trapping sets on the performance of iterative decoders.
In this context, our results can be seen as a generalization of
the findings in~\cite{schwartzetal06} for the case of trapping
sets.

An analytical study of the trapping redundancy is presented in the
following section.

\section{The Trapping Redundancy: A Probabilistic Approach} \label{sec:analytical_study}

We investigate the fundamental theoretical trade-offs between the
number of rows in a redundant parity-check matrix of a code and
the size of its smallest trapping set with a given set of
parameters.

\subsection{Definition and Bounds of the Trapping Redundancy}

In all our subsequent derivations, we make use of the following
definition.
\begin{definition} (\cite[p.\ 5]{hedayatetal99})\label{co-def}
An orthogonal array $\mathcal{A}$ of strength $t$ is an array of
dimensions $m \times n$ such that every $m \times t$ subarray
contains \emph{each} possible $t$-tuple as rows \emph{the same
number of times}.
\end{definition}

The codewords of an $[n,k,d]$ linear code $\mathcal{C}$ form an
orthogonal array of dimension $2^k \times n$ and strength
$d^{\perp}-1$, where $d^{\perp}$ denotes the dual distance of
$\mathcal{C}$.
Note that if $\mathcal{A}$ is an array of strength $t$, then
$\mathcal{A}$ is also an orthogonal array of strength $s$, for all
integers $s<t$.

Let $\theta_H(a,b)$ denote the number of $(a,b)$ trapping sets in
the parity-check matrix $\ve{H}$. We have the following result for
$\theta_H(a,b)$ corresponding to a matrix $\ve{H}$ that consists
of \emph{all} codewords of the dual code.

\begin{proposition} \label{prop1} Let $\ve{H}$ consist of all $2^{n-k}$
codewords of the dual code of an $[n,k,d]$ linear code
$\mathcal{C}$, for $n-k \geq 1$. Then $\theta_H(a,b)=0$ for all
pairs $(a,b)$ such that $1 \leq a \leq d-1$, and $b \neq
2^{n-k-1}$.
\end{proposition}

Proposition~\ref{prop1} shows that a parity-check matrix that
consists of all codewords of the dual code cannot contain trapping
sets with $1\leq a \leq d-1$ variables with less than or more than
$2^{n-k-1}$ checks connected to them an odd number of times. This
is a direct consequence of the fact that $\ve{H}$ in this case
represents an orthogonal array, so that each restriction of $1
\leq a \leq d-1$ columns of $\ve{H}$ contains each vector of
length $a$ the same number of times. Consequently, there are
$2^{n-k-1}$ rows in the restriction of the $a$ columns that have
even weight and $2^{n-k-1}$ rows that have odd weight.

However, it is of much larger importance to determine if there
exist parity-check matrices with a number of rows significantly
smaller than $2^{n-k}$ that are free of trapping sets with fixed
parameters $(a,s)$, for all $1 \leq s < b$. For this purpose, we introduce
the notion of the $(a,b)$ trapping redundancy of a code.
\begin{definition}
The $(a,b)$ \emph{trapping redundancy} $T_{a,b}(\mathcal{C})$ of
an $[n,k,d]$ linear code $\mathcal{C}$ is the smallest number of
rows $m$ of any (redundant) parity-check matrix which does not
contain trapping sets with parameters $(a,s)$, $1\leq s <b$.
Similarly, the smallest number of rows
$T_{a,b}^{(e)}(\mathcal{C})$ in a (redundant) parity-check matrix
of $\mathcal{C}$ avoiding $(a,s)$ elementary trapping sets with $1
\leq s < b$ is referred to as the $(a,b)$ \emph{elementary
trapping redundancy} of $\mathcal{C}$.
\end{definition}

\begin{theorem} \label{thm:trapp-imp}
Let $\mathcal{C}$ be an $[n,k,d]$ code and
$\mathcal{C}^\bot$ its dual. Use $\mathcal{M}_{\mathcal{C}}(m)$ to
denote the ensemble of all $m\times n$ matrices with rows chosen
independently and at random, with replacement, from the set of
$2^{n-k}$ codewords of $\mathcal{C}^\bot$. Furthermore, let $1
\leq a \leq \lfloor (d-1)/2 \rfloor$ be fixed, let $\Theta(a, b)$
be the number of trapping sets with parameters
$(a,s),\;0 \leq s < b$, 
$b \leq m$,  in a randomly chosen matrix from
$\mathcal{M}_{\mathcal{C}}(m)$, and let $\mathrm{e}$ denote the
base of the natural logarithm. If
\begin{equation} \label{eq:mi4}
{\mathrm{e}}\cdot \left( \binom{n}{a}-\binom{n-a}{a} \right) \, \left(\frac{1}{2}\right)^m\,
\sum_{j=0}^{b-1}\,\binom{m}{j} \leq 1,
\end{equation}
then $P\{{\Theta(a, b)=0\}}>0$. Consequently, if $m$ satisfies
\eqref{eq:mi4}, then there exists a parity-check matrix of
$\mathcal{C}$ with not more than $m+n-k-1$ rows
that does not contain any $(a,s)$ trapping sets with $1 \leq s < b$.
\end{theorem}

Note that $m+n-k-1$, for any $m$ satisfying~\eqref{eq:mi4},
represents an upper bound on the trapping redundancy of the code,
i.e.\, $T_{a,b}(\mathcal{C}) \leq m+n-k-1$.

\begin{proof} The proof of the claimed result is based on
Lov{\`a}sz Local Lemma (LLL), stated below following the
exposition of~\cite{alonetal00}.

\begin{lemma} Let $E_1,E_2,\ldots,E_N$ be a set of events in an
arbitrary probability space. Suppose that each event $E_i$ is
independent of all other events $E_j$, $1 \leq i,j \leq N$, except for at most $\tau$
of them, and that $P\{{E_i\}}\leq p$ for all $1 \leq i \leq N$. If
\begin{equation}
{\mathrm{e}}\;p\;(\tau+1) \leq 1,
\end{equation}
then $P\{{\bigcap_{i=1}^N\, \overline{E}_i\}}>0$.
\end{lemma}

Let $E_i$ be the event that the restriction of the $i$-th
collection (out of $\binom{n}{a}$) of $a$ columns from a randomly
chosen matrix in $\mathcal{M}_{\mathcal{C}}(m)$ contains fewer
than $b$ odd rows. Then
\begin{equation} \label{eq:PEi}
P\{ E_i \} = \left(\frac{1}{2}\right)^{m}\,
\sum_{j=0}^{b-1}\binom{m}{j}.
\end{equation}
Equation~\eqref{eq:PEi} follows from the fact that the codewords
of the dual code form an orthogonal array of strength $d-1$, and
that therefore even and odd weight rows in the restriction are
equally likely.
This is true independent of the choice of $a$, provided that $1
\leq a \leq \lfloor (d-1)/2 \rfloor$. The orthogonal array
property of strength $d-1$ implies that all row-vectors of length
up to $d-1$ are equally likely; henceforth, restricting the number
of columns $a$ to lie in the range $1\leq a \leq \lfloor (d-1)/2
\rfloor $ ensures that two events $E_i$ and $E_j$, $i \neq j$, are
independent as long as their corresponding sets of column indices
are disjoint.

In the above setting, $P\{{\bigcap\,\overline{E}_i\}}$ denotes the
probability that the randomly chosen matrix is free of trapping
sets with parameters $(a,s)$, for all $0 \leq s  < b$. 
In order to complete the proof, it suffices to observe that the following relationship holds for the dependence number $\tau$ of the events $E_i$,
\begin{equation}
\tau+1 = \sum_{l=1}^{a-1}\binom{a}{l}\,\binom{n-a}{a-l}+1
= \sum_{l=0}^{a}\binom{a}{l}\,\binom{n-a}{a-l}-\binom{n-a}{a}=\binom{n}{a}-\binom{n-a}{a}.
\label{eq:tau}
\end{equation}
Expression~\eqref{eq:tau} is a consequence of the fact that two
collections of $a$ columns are ``dependent'' if and only if they
share at least one column. In other words, if one collection of
$a$ columns is fixed, another collection of the same size is
deemed independent from it if its columns are chosen from the
remaining set of $n-a$ columns. The right-hand side of
Equation~\eqref{eq:tau} follows from the Vandermonde convolution
formula, which asserts that
\begin{equation}
\sum_l\;\binom{r}{t+l}\binom{s}{u-l}=\binom{r+s}{t+u}, \notag
\end{equation}
where $r,t,s,u$ denote non-negative integers.

In the worst case, additional $n-k-1$ rows may be needed to make
the randomly chosen matrix have full row-rank $n-k$.
This is due to the fact that a matrix containing no $(a,0)$ trapping sets must have rank at least one.

Note that appending $n-k-1$ additional rows to the selected set of
$m$ rows can only increase the number of odd-weight rows in each
restriction, and hence cannot reduce the value of the parameter
$b$. This completes the proof of the theorem.
\end{proof}

Although Theorem~\ref{thm:trapp-imp} ensures the existence of at
least one matrix in $\mathcal{M}_{\mathcal{C}}(m)$ that is free of
trapping sets with given parameters $(a,b)$, the actual
probability of selecting such a matrix may be very small. It is
therefore of interest to identify values of the parameter $m$ for
which the probability of drawing a matrix of the desired form
from the ensemble $\mathcal{M}_{\mathcal{C}}(m)$ is close to one.

\begin{theorem} \label{thm:trap_lll_high}
For a linear code $\mathcal{C}$, let
$\Theta(a,b)$ be the number of trapping sets with parameters
$(a,s)$, $1 \leq a \leq \lfloor (d-1)/2 \rfloor$,
$0 \leq s < b \leq m$, 
that exist in an arbitrarily chosen $m\times n$ array from the
$\mathcal{M}_{\mathcal{C}}(m)$ ensemble. If
\begin{equation} \label{eq:mi2}
\left( \frac{1}{2} \right)^{m} \sum\limits_{j=0}^{b-1} {m \choose
j} \leq \frac{\epsilon}{{n \choose a}} \left(1- \frac{\epsilon}{{n
\choose a}}\right)^{\tau},
\end{equation}
then $P\{{\Theta(a,b)=0\}}>1-\epsilon$, where $\tau$ is given by
Equation~\eqref{eq:tau}, and where $0<\epsilon<1$ is a real
number.

Consequently, if $m$ satisfies \eqref{eq:mi2}, then for small
values of $\epsilon$ one can find with high probability a
(redundant) parity-check matrix for $\mathcal{C}$ with not more
than $m+n-k-1$ rows
that does not contain any $(a,s)$ trapping sets with $1 \leq s < b$.

\end{theorem}
\begin{proof}
The result in Equation~\eqref{eq:mi2} is obtained from the
high-probability variation of LLL~\cite{alonetal00}, stated below.

\begin{lemma} Let $E_1,E_2,\ldots,E_{N}$ be a set of events
in an arbitrary probability space, and let $0<\epsilon<1$. Suppose
that each event $E_i$ is independent of all other events $E_j$,
except for at most $\tau$ of them. If
\begin{equation} \label{eq:p_Ei_hpv}
P\{{E_i\}}\leq
\frac{\epsilon}{N}\left(1-\frac{\epsilon}{N}\right)^{\tau},\; 1
\leq i \leq N,
\end{equation}
then $P\{{\bigcap_{i=1}^N\, \overline{E}_i\}}>1-\epsilon$.
\end{lemma}

To prove the theorem, let $E_{i}$, $1 \leq i \leq N$, denote the
event that the $i$-th collection of $a$ columns contains $0 \leq s < b$
rows of odd weight. Replace the expression for $P\{E_{i}\}$ in
Equation~\eqref{eq:p_Ei_hpv} by the right-hand side of
Equation~\eqref{eq:PEi} and use the formula for $\tau$ stated in
Equation~\eqref{eq:tau}.
\end{proof}

We derive next upper bounds on the elementary trapping redundancy
of linear block codes.

\begin{theorem} \label{thm:trapp-imp_elementary}
Let $\mathcal{C}$ be an $[n,k,d]$ code and $\mathcal{C}^\bot$ its
dual. Let $\mathcal{M}_{\mathcal{C}}(m)$ be the ensemble of all
$m\times n$ matrices with rows chosen independently and at random,
with replacement, from the set of $2^{n-k}$ codewords of
$\mathcal{C}^\bot$. Furthermore, let $1 \leq a \leq \lfloor
(d-1)/2 \rfloor$ be fixed and let $\Theta_e(a, b)$ be the number
of elementary trapping sets with parameters
$(a,s),\;0 \leq s < b,$ 
in a randomly chosen matrix of $\mathcal{M}_{\mathcal{C}}(m)$.
If
\begin{equation} \label{eq:trapping_redundancy_elementary}
{\mathrm{e}}\, \left( \binom{n}{a}-\binom{n-a}{a} \right) 
\frac{1}{2^{(a+1)\cdot m}}\cdot\sum_{j=0}^{b-1} \left[{m \choose j}
2^{j}a^{j} \cdot \left(a^2-a+2\right)^{m-j}\right] \leq 1,
\end{equation}
then $P\{{\Theta_e(a, b)=0\}}>0$. Consequently, if $m$ satisfies
\eqref{eq:trapping_redundancy_elementary}, then there exists a
parity-check matrix of $\mathcal{C}$ with no more than $m+n-k-1$
rows
that does not contain any $(a,s)$ elementary trapping sets with $1
\leq s < b$.
\end{theorem}

Note that $m+n-k-1$, with any $m$
satisfying~\eqref{eq:trapping_redundancy_elementary} represents an
upper bound on the elementary trapping redundancy, i.e.\
$T_{a,b}^{(e)} \leq m+n-k-1$.

\begin{proof} The proof follows the proof of
Theorem~\ref{thm:trapp-imp}. Let $E_i$ be the event that the
restriction of the $i$-th collection of $a$ columns from a
randomly chosen matrix in $\mathcal{M}_{\mathcal{C}}(m)$ contains
only rows of weight at most two. Among these rows, fewer than $b$
rows are required to have weight one.

Let $W_{\omega}$ denote the number of rows of weight $\omega$ in
the restriction of $a$ columns. Then
\begin{eqnarray} \label{eq:PEi_elementary}
\hspace{-0.5cm}P\{ E_i \}&\hspace{-0.3cm}=& \sum_{j=0}^{b-1} P\{ W_1 = j \; , \; (W_0 + W_2) = (m-j) \} \nonumber \\
 &\hspace{-0.3cm}=&  \sum_{j=0}^{b-1} \left[{m \choose j}\left(\frac{a}{2^a}\right)^{j} \cdot \left(\frac{a^2-a+2}{2^{a+1}}\right)^{m-j}\right] =  \frac{1}{2^{(a+1)\cdot m}} 
 \sum_{j=0}^{b-1} \left[{m \choose j}2^{j}a^{j} 
 \left(a^2-a+2\right)^{m-j}\right] ,
   \end{eqnarray}
where the second equation is a consequence of the fact that
\begin{equation}
P\{ W_1 = j \; , \; (W_0 + W_2) = (m-j)\}  = 
{m \choose
j} \left(\frac{\binom{a}{1}}{2^a}\right)^{j}\; \sum_{\ell=0}^{m-j}\,
\binom{m-j}{\ell}\,\left(\frac{1}{2^a}\right)^\ell\,\left(\frac{\binom{a}{2}}{2^a}\right)^{m-j-\ell}.
\notag
\end{equation}
Equation~\eqref{eq:PEi_elementary} follows from the observation
that the codewords of the dual code form an orthogonal array of
strength $d-1$, and that therefore all $1$-, $\binom{a}{1}$-, and
$\binom{a}{2}$-collections of rows of weight $0$, $1$, and $2$,
are equally likely, respectively.
\end{proof}

Similarly, based on the high-probability variation of LLL, one can
derive upper bounds on the number of rows needed to guarantee that
a randomly chosen matrix has no elementary trapping sets with a
given set of parameters with probability at least $1-\epsilon$.
The following theorem is an analogue of
Theorem~\ref{thm:trap_lll_high} for the case of elementary
trapping sets, i.e.\ with $P\{ E_i \}$ defined by
Equation~\eqref{eq:PEi_elementary}.

\begin{theorem} \label{thm:trap_lll_high_elementary}
For a linear code $\mathcal{C}$, let $\Theta_e(a,b)$ be the number
of elementary trapping sets with parameters $(a,s)$, $1 \leq a \leq \lfloor (d-1)/2 \rfloor$,
$1 \leq s < b \leq m$,
in an $m\times n$ array from the $\mathcal{M}_{\mathcal{C}}(m)$
ensemble. If

\begin{equation} \label{eq:trapping_redundancy_elementary_hp}
\frac{1}{2^{(a+1)\cdot m}} \sum\limits_{j=0}^{b-1} \left[{m\choose j}
2^{j}a^{j} \cdot \left(a^2-a+2\right)^{m-j}\right] \leq
\frac{\epsilon}{{n \choose a}} \left(1- \frac{\epsilon}{{n \choose
a}}\right)^{\tau},
\end{equation}

where $\tau$ is given in Equation~\eqref{eq:tau} and
$0<\epsilon<1$,
then $P\{{\Theta_e(a,b)=0\}}>1-\epsilon$.

For $m$ satisfying \eqref{eq:trapping_redundancy_elementary_hp}
and small values of $\epsilon$, every randomly constructed
parity-check matrix of $\mathcal{C}$ with no more than $m+n-k-1$
rows
does not contain $(a,s)$ elementary trapping sets with $1 \leq s <
b$.
\end{theorem}

If $E_{i}$, $1 \leq i \leq N$, is used to denote the event that
the $i$-th collection of $a$ columns contains only rows of weight
at
most two, and 
less than $b$ rows of weight one, then the result
represents a straightforward application of the high-probability
variation of LLL. The proof is therefore omitted.

\subsection{The Trapping Redundancy: A Constructive Approach}
\label{sec:constructive_study}

The problem of finding the trapping redundancy of a linear block
code can also be addressed in a deterministic manner, by invoking
arguments similar to those used for upper bounding the stopping
redundancy of codes. The results of this analysis are summarized
in the theorem below, for the case that the parameter $a$ of the
trapping set is bounded from above by $d-1$.

\begin{theorem} \label{th:constructive}
Let $\mathcal{C}$ be an $[n,k,d]$ linear code. Fix the parameter
$a \leq d-1$, and let $r=n-k>2$.\\ Then
\begin{equation} \label{eq:structured}
T_{a,b}(\mathcal{C}) \leq \sum_{i=1}^{t} \binom{r}{i} \notag
\end{equation}
where $b \geq 2^{a-1}-\sum_{j=t+1}^{a}\,\binom{r}{j}$ and $t \leq a$, \\
i.e.
\begin{equation}
T_{a,b}(\mathcal{C}) \leq \sum_{i=1}^{a}\, \binom{r}{i}+b-2^{a-1}.
\end{equation}
Furthermore, the smallest number of rows in a (redundant)
parity-check matrix avoiding elementary trapping sets with
parameters $(a,s)$, $s \leq b$, is upper bounded by
\begin{equation}
T_{a,b}^{(e)}(\mathcal{C}) \leq \sum_{i=1}^{a} \binom{r}{i}.
\end{equation}
\end{theorem}

Note that both claims in Theorem~\ref{th:constructive} also hold
for all trapping sets with parameters $(t,b)$, where $t \leq a$.

\begin{proof} Let $\ve{H}$ be an arbitrary parity-check matrix of the
code $\mathcal{C}$ of full row rank $r=n-k$. Consider the
restriction $\ve{H}_a$ of $\ve{H}$ on an arbitrary subset of $a$
of its columns. Due to the fact that $a \leq d-1$, these columns
are linearly independent, so that the row-rank of $\ve{H}_a$ must
be $a$. Hence, there exist $a$ rows in $\ve{H}_a$ that form a
basis for $\mathbb{F}_2^{a}$.

From $\ve{H}$, form a new redundant parity-check matrix $\ve{H}'$
by adding all linear combinations of at least two, but not more
than $t \leq a$ rows of $\ve{H}$, where $t$ will be determined
later. The total number of rows in $\ve{H}'$ in this case equals
the right hand side of the expression in
Equation~\eqref{eq:structured}. Since adding all possible linear
combinations of not more than $a$ rows of $\ve{H}$ to $\ve{H}'$
would ensure that $b \geq 2^{a-1}$, leaving out sums involving
$t+1,...,a$ rows can reduce the number of odd-weight rows in
$\ve{H}'_a$ by at most $\sum_{j=t+1}^{a}\, \binom{r}{j}$. This
completes the proof of the first claim. The proof of the second
claim represents a straightforward extension of the results
in~\cite{schwartzetal06}, and is therefore omitted.
\end{proof}

Simple inspection reveals that the bounds in
Equation~\eqref{eq:structured} are loose when compared to the
random bounds described in the previous section.

If $a\leq d-1$ and $b\leq r = n-k$, Equation~\eqref{eq:structured}
can be substantially tightened. In this case, it reads as
\begin{equation}
T_{(a,b)}(\mathcal{C}) \leq r+\binom{r}{2} = \frac{r (r+1)}{2},
\notag
\end{equation}
for general trapping sets, and
\begin{equation}
T_{(a,b)}(\mathcal{C}) \leq b\cdot r \notag
\end{equation}
for elementary trapping sets.

Let $\ell$ denote the number of odd-weight rows in an arbitrary
restriction $\ve{H}_a$ of $a$ columns on a parity-check matrix
$\ve{H}$. Since $\ve{H}$ has full rank, $\ve{H}_a$ contains at
least one odd-weight row and therefore $1 \leq \ell \leq r=n-k$
holds. It is now easy to show that adding all linear combinations
of two rows to the parity-check matrix $\ve{H}$ ensures that
$b\geq r$ holds for all $a$-sets of columns. For $1 \leq \ell \leq
r$, there are $\ell\cdot(r-\ell)$ odd-weight rows among all
$\binom{r}{2}$ linear combinations of pairs of rows. Adding those
linear combinations to the parity-check matrix brings the number
of odd-weight rows to $b = \ell + \ell\cdot(r-\ell) =
\ell\cdot(r-\ell+1) \geq r$. This follows from the simple
observation that the weight of the sum of one odd and one even
weight word is always odd. Therefore, to avoid $(a,b)$ trapping
sets with $a \leq d-1$, $b<n-k$, at most $r+\binom{r}{2} =
\frac{r(r+1)}{2}$ rows suffice.

As a final remark, note that in many applications, decoding
failure caused by trapping sets can only be detected upon
completion of the decoding process. In this case, one can choose
to add only a small number of judiciously chosen redundant rows to
the parity-check matrix of the code in order to eliminate the
influence of \emph{one particular trapping set}. How this can be
accomplished is illustrated on the example of the Margulis
$[2640,1320]$ code, in Section~\ref{app:Bounds_Margulis}.

\subsection{Asymptotic Formulas for the Trapping Redundancy}

Although there is no explicit formula for $m$ as defined by
Equations~\eqref{eq:mi4} and~\eqref{eq:mi2} that holds for all
possible parameter values $a$ and $b$, such a formula can be found
in the asymptotic regime ($m,n \to \infty$, $a=O(m)$, $b=O(m)$),
by using the following results from~\cite[p.~240]{hofri95}
and~\cite{brockwell64}.

Let
$$
A_{m} = \sum\limits_{0 \leq i \leq \lambda m} {m \choose i},
$$
where $0 \leq \lambda \leq 1$, and $b=\lfloor \lambda\,m \rfloor + 1$. Since small
values for the parameter $b$ are of special interest, assume that
$\lambda <1/2$. In this case we have
$$A_{m} \simeq {m \choose \lfloor \lambda m \rfloor} \cdot \frac{1}{1- \frac{\lambda}{1-\lambda}},$$
where the notation $c_m \simeq b_m$ describes the following
relationship between two functions $c_m$ and $b_m$ of $m$:
$\lim_{m \to \infty} c_m/b_m=1$.

For $b<m/2+1$, with $b = \lfloor \lambda m \rfloor +1$,
and $\lambda<1/2$, Equation~\eqref{eq:mi4} reduces to
\begin{equation}
\mathrm{e} \cdot \left( \binom{n}{a} - \binom{n-a}{a}\right) \cdot
\left(\frac{1}{2}\right)^{m} \cdot \binom{m}{\lfloor \lambda m
\rfloor} \cdot \frac{1}{1-\frac{\lambda}{1-\lambda}} \lesssim 1 .
\end{equation}
By invoking the well known asymptotic formula
\begin{equation}
\ld\, {m \choose \lfloor \lambda m \rfloor} \simeq m\,
{\mathrm{H}}_2\left(\frac{\lfloor \lambda m\rfloor}{m} \right),
\label{eq:log_shannon}
\end{equation}
where ${\mathrm{H}}_2(\cdot)$ denotes Shannon's binary entropy
function, and $\ld(\cdot)$ represents the logarithm with base two,
it follows that $m\leq m'$ with
\begin{equation}
m' \simeq \frac{\ld\left(\mathrm{e} \cdot \left( \binom{n}{a} - \binom{n-a}{a}\right) \right) + \ld\left( \frac{1}{1-\frac{\lambda}{1-\lambda}} \right)}{1-\mathrm{H}_2\left( \frac{\lfloor \lambda m\rfloor}{m} \right) }.
\end{equation}

Also, for $b<m/2+1$, with $b = \lfloor \lambda m \rfloor +1$, and
$\lambda<1/2$, the high-probability variation of LLL given in
Equation~\eqref{eq:mi2} reduces to
\begin{equation}
\left( \frac{1}{2} \right)^{m}\cdot  {m \choose \lfloor \lambda m
\rfloor} \cdot \frac{1}{1-\frac{\lambda}{1-\lambda}} \lesssim
\frac{\epsilon}{{n \choose a}} \cdot \left(1- \frac{\epsilon} {{n
\choose a}}\right)^{\left({n \choose a}-{n-a \choose a}-1\right)}.
\end{equation}

Using Equation~\eqref{eq:log_shannon} once again results in an
upper bound on the number of rows $m$ in a parity-check matrix
free of $(a,s)$ trapping sets, $0 \leq s < b$, i.e. $m\leq m'$ with

\begin{equation}
m' \simeq \frac{1}{H_2\left(\frac{\lfloor \lambda m\rfloor}{m} \right)-1}\; \left[\ld\left(1-\frac{\lambda}{1-\lambda}\right)+ \left(\binom{n}{a}-\binom{n-a}{a}-1\right)\, \ld\left(
\frac{\epsilon}{{n \choose a}} \left(1- \frac{\epsilon} {{n
\choose a}}\right)\right)\right].
\end{equation}

Note that for sufficiently large $m$ the right-hand side is not dependent on $m$ as $\frac{\lfloor \lambda m \rfloor}{m} \simeq \lambda$.  

\subsection{Asymptotic Formulas for the Elementary Trapping Redundancy}

Likewise, it is also of interest to find an explicit formula for
$m$ given by~\eqref{eq:trapping_redundancy_elementary} for the
case of elementary trapping sets. To this end, we use the
following asymptotic result taken from~\cite{brockwell64}, stating
that
\begin{equation}
\sum_{k=N}^{rN}\binom{rN}{k}\,p^k\;q^{rN-k} \simeq \phi(p^{-1})\;
\binom{rN}{N}\,p^N\,q^{rN-N},
\end{equation}
where $p,q>0$, $p+q=1$, $r>1$, $N$ is a positive integer, and
$\phi(y)=\frac{y-1}{y-r}$. By observing that the summands in
Equation~\eqref{eq:trapping_redundancy_elementary} can be
rewritten as
\begin{equation}
\sum_{j=0}^{b-1}\binom{m}{j}\,2^j\, a^j\, (a^2-a+2)^{m-j}=
(a^2+a+2)^m\sum_{u=m-b+1}^{m} \binom{m}{u}
\left(\frac{2a}{a^2+a+2}\right)^{m-u}\,
\left(\frac{a^2-a+2}{a^2+a+2}\right)^u, \notag
\end{equation}
it follows from substituting $m=rN$, $p = \frac{a^2-a+2}{a^2+a+2}$, and $q = \frac{2a}{a^2+a+2}$ that
\begin{equation}
\sum_{j=0}^{b-1}\binom{m}{j}\,2^j\, a^j\, (a^2-a+2)^{m-j} \simeq
\phi\left(\frac{a^2+a+2}{a^2-a+2}\right)\cdot 
\binom{m}{m-b+1}\left(\frac{2a}{a^2-a+2}\right)^{b-1}\;
\left(a^2-a+2\right)^m. \notag
\end{equation}
Using the above expression,
Equation~\eqref{eq:trapping_redundancy_elementary} reduces to
\begin{equation}
{\mathrm{e}} \cdot \left( \binom{n}{a}-\binom{n-a}{a} \right) \, \cdot
\phi\left(\frac{a^2+a+2}{a^2-a+2}\right)\;
\binom{m}{m-b+1}\left(\frac{2a}{a^2-a+2}\right)^{b-1}\;
\left(\frac{a^2-a+2}{2^{a+1}}\right)^m
 \lesssim 1. \notag
\end{equation}
This leads to a bound $m\leq m'$ with
\begin{eqnarray}
  m' \simeq \frac{ -\ld\left(  \mathrm{e} \left( \binom{n}{a}-\binom{n-a}{a}  \right) \cdot  \phi\left( \frac{a^2+a+2}{a^2-a+2}\right)  \right) \, - \,  (b-1)\cdot \ld\left( \frac{2a}{a^2-a+2}\right) } {\mathrm{H}_2\left( \frac{\lfloor \lambda m \rfloor}{m}\right) +   \ld(a^2-a+2)-(a+1)},
\end{eqnarray}
where we used Equation~\eqref{eq:log_shannon} to rewrite the right
hand side of the above expression. Similarly, for the
high-probability variation of LLL and for elementary trapping
sets, we obtain as a consequence of
Equation~\eqref{eq:trapping_redundancy_elementary_hp} the bound $m\leq m'$ with

\begin{equation}
 m' \simeq \frac{ \left(\binom{n}{a}-\binom{n-a}{a}-1\right) \cdot \ld\left( \frac{\epsilon}{{n \choose a}} \left(1- \frac{\epsilon}{{n \choose
a}}\right) \right)   -  
\ld\left( \phi\left( \frac{a^2+a+2}{a^2-a+2}\right) \right) - (b-1)\cdot \ld\left( \frac{2a}{a^2-a+2}\right)}
                            {\mathrm{H}_2\left( \frac{\lfloor \lambda m \rfloor}{m}\right) +   \ld(a^2-a+2)-(a+1)}\ .
\end{equation}

Based on the results of the previous sections, it is also
straightforward to see that the asymptotic formula for the
trapping redundancy of Theorem~\ref{th:constructive} is of the
form
\begin{equation}
T_{a,b}(\mathcal{C}) \lesssim \binom{r}{\lfloor \alpha \, r
\rfloor}\; \frac{1}{1-\frac{\alpha}{1-\alpha}}+b-2^{\lfloor \alpha\,r \rfloor -1},
\end{equation}
where $a=\lfloor \alpha\,r \rfloor$.

\textbf{Remark:} Note 
that the matrices from the ensemble
$\mathcal{M}_{\mathcal{C}}(m)$, for large $m$, may have highly
non-uniform row and column weights. The variable- and check-node
degrees of their corresponding Tanner graphs may be very large,
leading to the emergence of short cycles. It is therefore
important to compare the derived bounds with some benchmark
values, the latter corresponding to redundant matrices that are
known to have a small number of redundant rows, no short cycles,
as well as no small trapping sets. Two such examples, including
the aforementioned Margulis and projective geometry codes, are
discussed in the next section. Other families of codes, such as
codes based on Latin squares and designs, are analyzed in more
detail in the companion paper~\cite{laendneretal07}.

\section{Trapping Redundancy: Analytical Comparisons}
\label{sec:analytical_comparisons}

We perform next a numerical study of the probabilistic upper
bounds derived in Section~\ref{sec:analytical_study} for the
Margulis $[2640,1320]$ code and the class of projective geometry
codes. The goal of the comparative study is to both assess the
tightness of the bounds of Section~\ref{sec:analytical_study} and
to demonstrate that structured LDPC codes with redundant
parity-check matrices can avoid small trapping sets in their
Tanner graphs.

Note that the presented results only capture the trade-off between
the smallest size of general and elementary trapping sets and the
number of rows in the corresponding (redundant) parity-check
matrix, without taking into consideration other important matrix
properties such as variable and check nodes degree, girth, and
cycle length distribution.

\subsection{The $[2640,1320]$ Margulis Code} \label{app:Bounds_Margulis}

Numerical values of the trapping redundancy derived in
Section~\ref{sec:analytical_study} for the $[2640,1320]$ Margulis
code are listed in Tables~\ref{table:ts_bounds_margulis}
and~\ref{table:ts_bounds_margulis_elementary}. The labels
\emph{LLL (std)}, \emph{LLL  (hp)} refer to the bounds based on
LLL in standard form and its high-probability variation,
respectively. The symbol $m$ denotes the number of rows required
by the LLL approach, while $\hat{m}$ refers to the number of rows
of a redundant, rank $n-k$ parity-check matrix, which is an upper
bound on the trapping redundancy, $T_{a,b}(\mathcal{C}) \leq
\hat{m}$. As the exact minimum distance is not known for this
code, a method~\cite{hu} for approximating the minimum distance,
proposed in~\cite{huetal04}, was used instead. The estimate at
hand is $d\approx 40$, and we restrict our attention to values of
the parameter $a$ strictly (and significantly) smaller than
$a/2=20$.

The full-rank $1320\times 2640$ parity-check matrix $\ve{H}$ of
the Margulis code, constructed in the standard
manner~\cite{margulis82}, contains no cycles of length less than
eight, but includes a large number of $(12,4)$ and $(14,4)$
elementary trapping sets~\cite{richardson03,laendneretal05}. The
LLL-based bounds reveal that there exists a matrix of full rank
with at most $1336$ rows that does not contain $(14,s)$ elementary
trapping sets, with $s < 5$, and that adding at most $19$
additional rows ensures that the matrix does not contain $(12,4)$
trapping sets either.

It can also be seen from Table~\ref{table:ts_bounds_margulis} that
there exists a matrix of row-rank $n-k$ with $\hat{m}=1394$ rows
that is free of trapping sets of size $(6,s)$, $s<5$. However, for
a parity-check matrix free of elementary trapping sets of the same
parameters, Table~\ref{table:ts_bounds_margulis_elementary} shows
that only $\hat{m}=1351$ rows are needed.

These bounds cover the case of fixed $a$ values only. Note that
finding analogs of these results that cover a \emph{range} of
general trapping set sizes $a$ instead may be desirable for
certain practical applications. Due to the monotonic increase of
the trapping redundancy with the value of the parameter $a$, one
can see that if a set of rows, randomly drawn from the codewords
of the dual code, does not contain trapping sets of size $a$ with
high probability, then this set is also very unlikely to support
trapping sets of size smaller than $a$. For example, to obtain a
matrix free of $(14,s)$ trapping sets, $s<5$, according to the
high-probability version of LLL with $\epsilon = 10^{-20}$ one
needs $m=216$ rows, so that $\hat{m}=1535$. With \emph{high
probability}, this matrix does also avoid $(12,s)$ trapping sets,
$s<5$, due to the LLL-based study.

Note that for elementary trapping sets,
Table~\ref{table:ts_bounds_margulis} indicates that the larger the
value of the parameter $a$, the smaller the number of redundant
rows that is needed to eliminate such trapping sets. This result
may seem counterintuitive, but it follows from the fact that
trapping sets are deemed elementary only as their restriction does
not contain rows of weight larger than two - an event that becomes
less likely with the increase of the parameter $a$.

\begin{table}[htdp]
\begin{center}\begin{tabular}{|cc|c|p{.5cm}p{.5cm}|p{.5cm}p{.5cm}|p{.5cm}p{.5cm}|p{1cm}p{1cm}|}
 \hline
 \multicolumn{2}{|c|}{TS size} & $\ve{H}$  & \multicolumn{2}{c|}{LLL (std)}&
 \multicolumn{2}{c|}{LLL (hp)} &\multicolumn{2}{c|}{LLL (hp)} \\
        &         &                              &          &
        &  \multicolumn{2}{c|}{$\epsilon = 0.01$}&\multicolumn{2}{c|}{$\epsilon = 10^{-20}$}\\
$a$ & $b$ & $(n-k) \times n$ & $m$ & $\hat{m}$ & $m$ & $\hat{m}$ & $m$ & $\hat{m}$ \\
\hline
$6$ & $5$ & $1320\times 2640$ &  $~~75$ & $1394$ & $~~87$ & $1406$ & $150 $ & $1469 $\\
$8$ & $5$ & $1320\times 2640$ &  $~~94$ & $1413$ & $105$ & $1424$ & $167 $ & $1486 $\\
$12$ & $5$ & $1320\times 2640$ &  $129$ & $1448$ & $138$ & $1457$ & $200 $ & $1519 $\\
$14$ & $5$ & $1320\times 2640$ &  $145$ & $1464$ & $154$ & $1473$ & $216$ & $1535$\\
\hline\end{tabular} \caption{\label{table:ts_bounds_margulis}
Upper bounds on the $(a,b)$ trapping redundancy $T_{a,b}(\mathcal{C})$ of the Margulis code.}
\end{center}
\end{table}

\begin{table}[htdp]
\begin{center}\begin{tabular}{|cc|c|cc|cc|cc|cc|}
 \hline
 \multicolumn{2}{|c|}{TS size} & $\ve{H}$  & \multicolumn{2}{c|}{LLL (std)}&
 \multicolumn{2}{c|}{LLL (hp)} &\multicolumn{2}{c|}{LLL (hp)} \\
        &         &                              &          &                   &  \multicolumn{2}{c|}{$\epsilon = 0.01$}&\multicolumn{2}{c|}{$\epsilon = 10^{-20}$}\\
$a$ & $b$ & $(n-k) \times n$ & $m$ & $\hat{m}$ & $m$ & $\hat{m}$ & $m$ & $\hat{m}$ \\
\hline
$6$ & $5$ & $1320\times 2640$ & $32$ & $1351$ & $39$ & $1358$ & $70$ & $1389$\\
$8$ & $5$ & $1320\times 2640$ &  $26$ & $1345$ & $29$ & $1348$ & $49$ & $1368$  \\
$12$ & $5$ & $1320\times 2640$ &  $19$ & $1338$ & $20$ & $1339$ & $31$ & $1350$  \\
$14$ & $5$ & $1320\times 2640$ & $17$ & $1336$ & $18$ & $1337$ & $26$ & $1345$\\\hline
\end{tabular} \caption{
\label{table:ts_bounds_margulis_elementary}
Upper bounds on the $(a,b)$ elementary trapping redundancy $T_{a,b}^{(e)}(\mathcal{C})$ of the Margulis code.}
\end{center}
\end{table}

We complete this section by illustrating how a simple
\emph{structured method} can be used to add only one redundant
parity-check equation so as to increase the value of the parameter
$b$ in any given $({a=12},{b=4})$ or $({a=14},{b=4})$ trapping set.

\begin{example}
The knowledge about the structure of specific trapping sets - such
as the $(12, 4)$ and $(14, 4)$ trapping sets - in the $[2640,
1320]$ Margulis code can be used to eliminate single instances of
such sets.

First, we observe that the support of any $(14,4)$ trapping set contains
 the support of a $(12,4)$ trapping set - i.e., the sets
are nested. Throughout the remainder of this section, we call the
variables and checks introduced by extending a $(12, 4)$ to a
$(14, 4)$ trapping set \emph{expansion variables} and
\emph{expansion checks}, respectively. The notation
${\mathrm{B}}$, ${\mathrm{E}}$, and ${\mathrm{O}}$ is used to
refer to the {\underline{b}}asic $(12, 4)$ trapping set, its
{\underline{e}}xpansion variables and checks, and the graph
{\underline{o}}utside (i.e. complementary to) the $(14, 4)$
trapping set, respectively. Since the Margulis code is regular,
with variable node degree $d_v=3$ and check node degree $d_c=6$,
an elementary $(a, b)$ trapping set with $a$ variables and $b$
check nodes of degree one\footnote{These check nodes,
incidentally, correspond to unsatisfied check nodes, that can be
readily identified upon termination of the iterative decoding
process.} has a fixed number of checks. Therefore, in order to
extend an elementary $(12, 4)$ trapping set to a $(14, 4)$
trapping set, two variables and three checks have to be added. The
notions of basic, expansion and outside variables and check
nodes are illustrated in Figure~\ref{fig:trapping_set_structure} 

Since the Margulis code has no four-cycles, at most one expansion
check can be connected to both the expansion variables.
Consequently, either three or four edges emanating from the
expansion variables are connected to expansion check nodes, while
the remaining edges are connected to check nodes whose degree in
the basic $(12, 4)$ trapping set is one. Thus there exist only two
possible configurations for such trapping sets, as shown in
Figure~\ref{fig:trapping_set_structure}.

Now, based only on the knowledge of check nodes of degree one
within the basic $(12, 4)$ trapping set, it is straightforward to
determine the whole expansion set: in the first case, the two
degree-one checks of the basic $(12, 4)$ trapping set are
connected through two variable nodes to one additional check node.
These two variable nodes are the expansion variables. In the
second configuration, the expansion variables do not share a check
node.

\begin{figure}
\begin{center}
\psfrag{VBO1}[r][r]{\footnotesize{ $v_{\subBO,1}$ }}
\psfrag{VBO2}[r][r]{\footnotesize{ $v_{\subBO,2}$ }}
\psfrag{VBE1}[r][r]{\footnotesize{ $v_{\subBE,1}$ }}
\psfrag{VBE2}[r][r]{\footnotesize{ $v_{\subBE,2}$ }}
\psfrag{VE1}[r][r]{\footnotesize{ $v_{\subE,1}$ }}
\psfrag{VE2}[r][r]{\footnotesize{ $v_{\subE,2}$ }}
\psfrag{CBO1}[l][l]{\footnotesize{ $c_{\subBO,1}$ }}
\psfrag{CBO2}[l][l]{\footnotesize{ $c_{\subBO,2}$ }}
\psfrag{CBE1}[l][l]{\footnotesize{ $c_{\subBE,1}$ }}
\psfrag{CBE2}[l][l]{\footnotesize{ $c_{\subBE,2}$ }}
\psfrag{CE}[l][l]{\footnotesize{ $c_{\subE}$ }}
\psfrag{CEO1}[l][l]{\footnotesize{ $c_{\subEO,1}$ }}
\psfrag{CEO2}[l][l]{\footnotesize{ $c_{\subEO,2}$ }}
\psfrag{CEO3}[l][l]{\footnotesize{ $c_{\subEO,3}$ }}
\psfrag{CBE3}[l][l]{\footnotesize{ $c_{\subBE,3}$ }}
\psfrag{VBE3}[r][r]{\footnotesize{ $v_{\subBE,3}$ }}
\psfrag{TS}[l][l]{\tiny{(12, 4) trapping set}}
\psfrag{EXT}[l][l]{\tiny{Expansion set}}
\psfrag{BIGTS}[l][l]{\tiny{(14, 4) trapping set}}
\psfrag{A}[c][c]{\small{(a)}} \psfrag{B}[c][c]{\small{(b)}}
\includegraphics[scale=0.24]{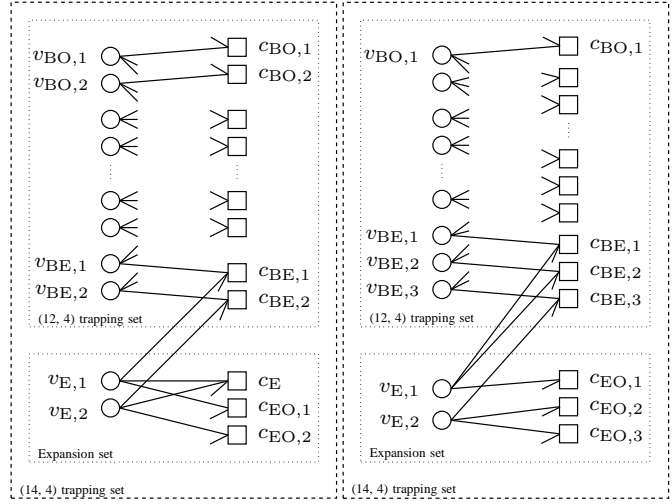}
\caption{Trapping set structure of a $(12, 4)$ trapping set and
its
  expansion. (a) first configuration; (b) second configuration.}
\label{fig:trapping_set_structure}
\end{center}
\end{figure}

For the first configuration, denote the two expansion variables by
$v_{\subE,1}$ and $v_{\subE,2}$. Furthermore, denote the check
node connected to both these variables by $c_{\subE}$. The
degree-one check nodes in the basic trapping set neighboring the
expansion variables are denoted by $c_{\subBE,1}$ and
$c_{\subBE,2}$, while the variable nodes in the basic trapping set
connected to check nodes $c_{\subBE,1}$ and $c_{\subBE,2}$ are
denoted by $v_{\subBE,1}$ and $v_{\subBE,2}$, respectively. The
check nodes of degree one in the expansion of the trapping set are
named $c_{\subEO,1}$ and $c_{\subEO,2}$. The two remaining check
nodes of degree one in the basic trapping set are termed
$c_{\subBO,1}$ and $c_{\subBO,2}$, and the variable nodes within
the basic $(12, 4)$ trapping set connected to them are $v_{\subBO,1}$
and $v_{\subBO,2}$, respectively (see
Figure~\ref{fig:trapping_set_structure}(a)). The configuration
involving all the aforementioned checks and variables is
illustrated in Table~\ref{table:structured_row_combining}.

\begin{table}[htdp]
\begin{center}
\begin{tabular}{l|cc|cccc|}
& \multicolumn{2}{c}{Expansion Variables} & \multicolumn{4}{c}{Basic Trapping Set Variables} \\
& $v_{\subE,1}$ & $v_{\subE,2}$ & $v_{\subBE,1}$ & $v_{\subBE,2}$ & $v_{\subBO,1}$ & $v_{\subBO,2}$ \\
\hline
$c_{\subE}$       & 1 & 1 & 0 & 0 & 0 & 0 \\
$c_{\subEO,1}$ & 1 & 0 & 0 & 0 & 0 & 0 \\
$c_{\subEO,2}$ & 0 & 1 & 0 & 0 & 0 & 0 \\
$c_{\subBE,1}$ & 1 & 0 & 1 & 0 & 0 & 0 \\
$c_{\subBE,2}$ & 0 & 1 & 0 & 1 & 0 & 0 \\
$c_{\subBO,1}$ & 0 & 0 & 0 & 0 & 1 & 0 \\
$c_{\subBO,2}$ & 0 & 0 & 0 & 0 & 0 & 1
\end{tabular}
\end{center}
\caption{Restriction of the basic trapping set and the expansion
variables} \label{table:structured_row_combining}
\end{table}%

The parity-check equations containing the restriction described in
Table~\ref{table:structured_row_combining} can be linearly
combined to generate a redundant parity-check equation that has a
restriction of odd weight in both the $(12, 4)$ and the $(14, 4)$
trapping set. There are three different methods for linearly
combining the parity-check equations with restrictions as shown in
Table~\ref{table:structured_row_combining}.

Method {\bf $S_1$} refers to adding the rows indexed by
($c_{\subE}$, $c_{\subEO,1}$, $c_{\subBE,2}$),\ ($c_{\subE}$,
$c_{\subEO,2}$, $c_{\subBE,1}$),\ ($c_{\subEO,1}$,
$c_{\subBE,1}$), and ($c_{\subEO,2}$, $c_{\subBE,2}$). Method {\bf
$S_2$} refers to adding the rows indexed by ($c_{\subE}$,
$c_{\subEO,1}$) and ($c_{\subE}$, $c_{\subEO,2}$). Observe that
all these combinations have a restriction of weight one within the
expansion variables. Method {\bf $S_3$} differs from the previous
methods in so far that it may generate redundant parity-check
equations with odd-weight restrictions that are not necessarily of
weight one. Candidate equations are obtained by adding the rows
indexed by ($c_{\subE}$, $c_{\subBO,1}$), ($c_{\subE}$,
$c_{\subBO,2}$),\ ($c_{\subE}$, $c_{\subBE,1}$, $c_{\subBE,2}$,
$c_{\subBO,1}$),\ ($c_{\subE}$, $c_{\subBE,1}$, $c_{\subBE,2}$,
$c_{\subBO,2}$),\ ($c_{\subBO,1}$, $c_{\subBO,2}$, $c_{\subEO,1}$,
$c_{\subBE,2}$), and ($c_{\subBO,1}$, $c_{\subBO,2}$,
$c_{\subEO,2}$, $c_{\subBE,1}$). The Hamming weight of the
constructed rows is an even integer between $10$ and $24$. The
lower bound $10$ is obtained if the supports of two added
parity-checks share exactly one element. The intersection of the
supports cannot have more than one element, since the code has no
four cycles. The upper bound $24$ is met when four parity-check
equations are added and none of the variables listed in Table
\ref{table:structured_row_combining} occur in the support of more
than one parity-check.

If a trapping set of the form shown in
Figure~\ref{fig:trapping_set_structure}(b) is present in the code
graph, then the two expansion variables cannot have a common check
node. Consequently, the expansion variable $v_{\subE,1}$ is
connected to two of the degree-one check nodes of the basic
trapping set.\footnote{One must keep in mind that the choice for
such a variable may not be unique.} If there is only one variable
node connected to two degree-one checks, denoted by $c_{\subBE,1}$
and $c_{\subBE,2}$, the variable of interest is the expansion
variable $v_{\subE,1}$, which is also connected to expansion check
node $c_{\subEO,1}$. Observe that the expansion variable
$v_{\subE,2}$ is strongly influenced by its two neighboring check
nodes connected to the outside graph and it cannot be uniquely
determined. Due to this limited knowledge of the expansion set,
there are only two possible ways to generate a redundant
parity-check with an odd restriction weight on the basic trapping
set, involving the sums of the rows ($c_{\subEO,1}$,
$c_{\subBE,1}$) as well as ($c_{\subEO,1}$, $c_{\subBE,2}$). A
similar analysis can be conducted for $(14,4)$ trapping sets.
Details regarding this procedure are omitted.
\end{example}

As illustrated by the example, knowledge about the structure of
trapping sets allows one to exactly determine the choice of the
redundant row to be added to the parity check matrix.
Unfortunately, since there are at least $1320$ trapping sets of
each such form in the code graph, adding this many rows to the
code matrix is undesirable. Nevertheless, as already pointed out,
only one row can be added upon detecting the presence of a
decoding failure caused by a given trapping set.

\subsection{Projective Geometry Codes} \label{sec:pg}

Projective geometry codes are linear block codes with many well
known combinatorial parameters and properties. As will be shown
next, it is also straightforward to characterize a large
sub-family of trapping sets in these codes.

We start our derivations by introducing the relevant terminology.

\begin{definition}~\cite{macwilliamsetal77} A finite projective
geometry $\PG(M,q)$ of dimension $M$, over a finite field
$\GField(q)$, for some prime power $q$, is a set of points and
subsets thereof, called lines. The following axioms hold for the
points and lines of a finite geometry:
\begin{itemize}
\item Two distinct points determine a unique line. \item Every
line consists of more than two points. \item For every pair of
distinct lines $L_1$ and $L_2$, intersecting at some point $r$,
there exist two pairs of points $(p_1,q_1) \in L_1$ and $(p_2,q_2)
\in L_2$ that differ from $r$, such that the lines determined by
$(p_1,p_2)$ and $(q_1,q_2)$ intersect as well. \item For each
point and for each line, there exist at least two lines and two
points that are not incident to them, respectively.
\end{itemize}
\end{definition}
The points of a projective geometry $\PG(M,q)$ can be represented
by non-zero $(M+1)$-tuples $(a_0,a_1,a_2,...,a_M)$ such that $a_i
\in \GField(q)$. Points of the form $(a_0,a_1,a_2,...,a_M)$ and
$(\delta a_0,\delta a_1,\delta a_2,...,\delta a_M)$, $\delta \in
\GField(q) \backslash \{{0\}}$, are considered equivalent. A line
through two distinct points $(a_0,a_1,a_2,...,a_M)$ and
$(b_0,b_1,b_2,...,b_M)$ consists of all points that can be
expressed as
$(\alpha\,a_0+\beta\,b_0,...,\alpha\,a_M+\beta\,b_M)$, where
$\alpha,\beta \in \GField(q)$ and are not both simultaneously
zero. Consequently, a projective geometry $\PG(M,q)$ has
$(q^{M+1}-1)/(q-1)$ points, and each line in the geometry contains
$q+1$ points. The number of lines in a projective geometry is
given by
\begin{equation}
(q^M+...+q+1)(q^{M-1}+...+q+1)/(q+1). \label{eq:number_of_lines}
\end{equation}

It is straightforward to see that the number of lines and points
coincide for $M=2$, since $(q^2+q+1)={(q^{3}-1)}/{(q-1)}$ holds.

A type-I projective geometry code is defined in terms of a
parity-check matrix representing the \emph{line-point} incidence
matrix of a projective geometry $\PG(M,q)$~\cite{kouetal01}.
Throughout the remainder of the paper, we consider projective
plane codes, $M=2$, and codes based on projective geometries with
$M=3$ only.
\begin{definition}
An $s$-arc in $\PG(2,q)$ is a collection of $s$ points such that no
three of them are collinear. The lines incident to an $s$-arc
$\mathcal{K}$ are either unisecants (they intersect the arc in
exactly one point) or bisecants (they intersect the arc in exactly
two points). Similarly, an $s$-cap in $\PG(3,q)$ is a set of $s$
points, no three of which are collinear.
\end{definition}
The following results pertaining to unisecants and bisecants are
taken from~\cite[Ch. 8]{hirschfeld79} and~\cite[Ch.
16]{hirschfeld85}.
\begin{lemma} \label{lemma:secants} Let $n_1$ and $n_2$ denote the number of unisecants
and bisecants of an $s$-arc $\mathcal{K}$ in $\PG(2,q)$, respectively. Then
\begin{equation}
n_1=s(q+2-s),\;\;\text{and}\;\;\; n_2=\frac{1}{2}s(s-1).
\end{equation}
Similarly, for an $s$-cap $\mathcal{K}$ in $\PG(3,q)$ it holds
that
\begin{equation}
n_1=s(q^2+q+2-s),
\end{equation}
where $n_1$ denotes the number of unisecants of $\mathcal{K}$.
\end{lemma}
\begin{lemma} \label{lemma:arcs} The largest arc in $\PG(2,q)$ contains at most $q+2$
points, for $q$ even, and $q+1$ points, for $q$ odd. Arcs with
$s=q+1$ and $s=q+2$ are called ovals and hyperovals, respectively.
The size of any $s$-cap in $\PG(3,q)$ satisfies $s \leq q^2+1$. For
$q>2$, a $(q^2+1)$-cap is called an ovaloid.
\end{lemma}

The results of Lemmas~\ref{lemma:secants} and~\ref{lemma:arcs} can
be used to establish the following simple results regarding
trapping sets in the Tanner graph of type-I projective geometry
codes. Note that all stated results restrict the parameter sets
for which trapping sets may potentially exist, although they do
not imply the existence of such sets.

\begin{corollary}
\label{cor:pg_to_code_one} All elementary trapping sets of a
$\PG(2,q)$, type-I, projective geometry code have parameters
$(s,s(q+2-s))$. Consequently, the number of degree-one check nodes
of such trapping sets for $q$ odd is necessarily larger than or
equal to the number of variables in the trapping set. For even
values of $q$, an exception to the aforementioned rule is a
hyperoval, which represents a codeword. The trapping sets with the
smallest ratio $b/a$ have parameters $(q+1,q+1)$ ($q$ odd) and
$(q+2,0)$ ($q$ even), respectively, and those with the largest
ratio $(3,3(q-1))$.
\end{corollary}

\begin{proof}
Note that the parity-check matrix $\ve{H}$ of a $\PG(2,q)$ code is
the line-point incidence matrix of the underlying $\PG$. Arcs
correspond to a collection of columns, the restriction of which
has rows of weight at most two only, and exactly $n_1$ of these
rows have weight one. This is equivalent to the definition of an
$(s,n_1)$ trapping set, where $s \geq 3$ and also $s \leq (q+1)$
($q$ odd) or $s \leq (q+2)$ ($q$ even), respectively, according to
Lemma~\ref{lemma:arcs}. The smallest and largest ratios of $b/a$
are defined by the limits of $s$. Hyperovals have $s=q+2$ points
and $n_1 = (q+2)\cdot(q+2-(q+2))=0$ unisecants, and therefore
correspond to $(q+2,0)$ trapping sets, which are codewords.
\end{proof}

\begin{corollary}
\label{cor:pg_to_code_two} All elementary trapping sets of a
$\PG(3,q)$, type-I projective geometry code have parameters
$(s,s(q^2+q+2-s))$. Provided that the PG contains an $s$-arc with
$n_1=s(q^2+q+2-s)$, the trapping sets with the smallest and
largest ratio $b/a$ have parameters $(q^2+1,(q^2+1)(q+1))$ and
$(3,3(q^2+q-1))$, respectively.
\end{corollary}

\begin{proof}
The proof follows along the lines of the proof of
Corollary~\ref{cor:pg_to_code_one}, with $n_1 = s(q^2+q+2-s)$ for
$\PG(3,q)$.
\end{proof}

A complete classification of trapping sets
in projective geometry codes is probably an impossible task. This
is due to the fact that very little is known about the number and
existence of arcs and caps of different sizes in projective
spaces. One aspect of this problem that is better understood is
the existence and enumeration of \emph{complete arcs} (and
\emph{caps}) - i.e. arcs and caps not contained in any larger arc
or cap. The interested reader is referred to~\cite{hirschfeld79,hirschfeld85} for more
information regarding the problem of complete arc enumeration.

We compare next the upper bounds derived in
Section~\ref{sec:analytical_study} with the results of the study
presented in this section. We consider elementary trapping sets
only.

In order to apply the results of the lemmas in this section, one
has to consider a parity-check matrix that represents a complete
line-point incidence structure~\cite{kouetal01}. For this reason,
the standard parity-check matrices of PG codes contain exactly $n$
rows, and are therefore redundant.

Table~\ref{table:ts_bounds_pg_elementary} list the number of rows
required to avoid elementary trapping sets of a given size,
computed according to LLL and its high-probability variation. The
values for $\hat{m}$ are derived using the minimum distance and
the rank results taken from~\cite{kouetal01}.

Observe that Table~\ref{table:ts_bounds_pg_elementary} indicates
that there exists a parity-check matrix for the $\PG(2,16)$ code
with at most $175$ rows and no $(3,s)$, $s<45$, elementary
trapping sets. This is significantly less than $n=273$
 as for the $\PG$ code, but might also include rows of large weight.
On the other hand, to obtain a matrix without such trapping sets
with probability larger than $1-10^{-20}$, at most $309$ rows are
required, a number clearly larger than $n$.

\begin{table*}[!th]
\begin{center}\begin{tabular}{|c|cc|c|c|cc|cc|cc|}
 \hline
 Code &\multicolumn{2}{c|}{TS size} & compl. $\ve{H}$ & min. $\ve{H}$  & \multicolumn{2}{c|}{LLL (std)} & \multicolumn{2}{c|}{LLL (hp)} &\multicolumn{2}{c|}{LLL (hp)} \\
& & & & & & &  \multicolumn{2}{c|}{$\epsilon = 0.01$}&\multicolumn{2}{c|}{$\epsilon = 10^{-20}$}\\
 &  $a$ & $b$ & $\overline{m}=n$ & ${\underline{m}}=n-k$ & $m$ & $\hat{m}$ & $m$ & $\hat{m}$ & $m$ & $\hat{m}$ \\
\hline
$\PG(2,16)$ & $3$ & $45$ & $273$ & $82$ & $94$ & $175$ & $125$ & $206$ & $228$ & $309$  \\
$\PG(2,16)$ & $8$ & $80$ & $273$ & $82$ & $80$ & $161$ & $80$ & $161$ & $80$ & $161$  \\
$\PG(2,32)$ & $3$ & $93$ & $1057$ & $244$ & $115$ & $358$ & $178$ & $421$ & $338$ & $581$  \\
$\PG(2,32)$ & $16$ & $288$ & $1057$ & $244$ & $288$ & $531$ & $288$ & $531$ & $288$ & $531$\\
\hline\end{tabular} \caption{
Upper bounds on the $(a,b)$ elementary trapping redundancy of PG codes.
\label{table:ts_bounds_pg_elementary}}
\end{center}
\end{table*}

Although there exist parity-check matrices for the $\PG$ codes
that have smaller row-redundancy and do not contain elementary
trapping sets of small sizes, the matrix defined by the $\PG$
construction presents a way to generate a parity-check matrix with
limited redundancy, small row-weight, and therefore few short
cycles, which also performs well in the waterfall region. As can
be seen from the comparison table, $\PG$ codes represent an
attractive systematic construction of LDPC codes without small
trapping sets, whose redundancy lies between the bounds based on
the standard case and the high-probability version of LLL.

\section{Conclusions}\label{sec:conclusions}

We introduced the notion of the $(a,b)$ trapping redundancy of a
code, representing the smallest number of rows in any parity-check
matrix of the code that avoids $(a,s)$ trapping sets with $1\leq s
< b$. Upper bounds on these combinatorial numbers were derived
using Lov{\`a}sz  Local Lemma and variations thereof. Also
presented were numerical results for the trapping redundancy of
the Margulis $[2640,1320]$ and type-I PG codes.

\vspace{3mm}

\textbf{Acknowledgment:} The authors are grateful to the anonymous
reviewers for their constructive comments that significantly
improved the exposition of the work. In addition, they would like
to thank Dr. Richardson for carefully handling the manuscript.


\begin{thebibliography}{10}
\providecommand{\url}[1]{#1}
\csname url@rmstyle\endcsname
\providecommand{\newblock}{\relax}
\providecommand{\bibinfo}[2]{#2}
\providecommand\BIBentrySTDinterwordspacing{\spaceskip=0pt\relax}
\providecommand\BIBentryALTinterwordstretchfactor{4}
\providecommand\BIBentryALTinterwordspacing{\spaceskip=\fontdimen2\font plus
\BIBentryALTinterwordstretchfactor\fontdimen3\font minus
  \fontdimen4\font\relax}
\providecommand\BIBforeignlanguage[2]{{%
\expandafter\ifx\csname l@#1\endcsname\relax
\typeout{** WARNING: IEEEtran.bst: No hyphenation pattern has been}%
\typeout{** loaded for the language `#1'. Using the pattern for}%
\typeout{** the default language instead.}%
\else
\language=\csname l@#1\endcsname
\fi
#2}}

\bibitem{dietal02}
C.~Di, D.~Proietti, I.~Telatar, T.~Richardson, and R.~Urbanke, ``Finite-length
  analysis of low-density parity-check codes on the binary erasure channel,''
  \emph{IEEE Transactions on Information Theory}, vol.~48, no.~6, pp.
  1570--1579, June 2002.

\bibitem{mackayetal03}
\BIBentryALTinterwordspacing
D.~MacKay and M.~Postol, ``Weaknesses of {M}argulis and {R}amanujan-{M}argulis
  low-density parity-check codes,'' \emph{Electronic Notes in Theoretical
  Computer Science}, vol.~74, 2003. [Online]. Available:
  \url{http://www.cs.toronto.edu/~mackay/margulis.pdf}
\BIBentrySTDinterwordspacing

\bibitem{richardson03}
T.~Richardson, ``Error-floors of {LDPC} codes,'' in \emph{Proceedings of the
  41st Annual Allerton Conference on Communication, Control and Computing},
  Monticallo, Illinois, USA, September 2003, pp. 1426--1435.

\bibitem{hanetal07}
J.~Han and P.~Siegel, ``Improved upper bounds on stopping redundancy,''
  \emph{IEEE Transactions on Information Theory}, vol.~53, no.~1, pp. 90--104,
  January 2007.

\bibitem{koetteretal06}
R.~Koetter, W.-C.~W. Li, P.~Vontobel, and J.~Walker, ``Characterizations of
  pseudo-codewords of {LDPC} codes,'' \emph{accepted for Advances in
  Mathematics}, August 2006.

\bibitem{schwartzetal06}
M.~Schwartz and A.~Vardy, ``On the stopping distance and stopping redundancy of
  codes,'' \emph{IEEE Transactions on Information Theory}, vol.~52, no.~3, pp.
  922--932, March 2006.

\bibitem{weberetal05}
J.~Weber and K.~Abdel-Ghaffar, ``Stopping and dead-end set enumerators for
  binary {H}amming codes,'' in \emph{Proceedings of the Twenty-sixth Symposium
  on Information Theory in the Benelux}, Brussels, Belgium, May 2005, pp.
  165--172.

\bibitem{hedayatetal99}
A.~Hedayat, N.~Sloane, and J.~Stufken, \emph{Orthogonal Arrays: Theory and
  Applications}.\hskip 1em plus 0.5em minus 0.4em\relax New York, USA: Springer
  Verlag, 1999.

\bibitem{milenkovicetal06}
O.~Milenkovic, E.~Soljanin, and P.~Whiting, ``Stopping and trapping sets in
  generalized covering arrays,'' in \emph{Proceedings of the 40th annual
  Conference on Information Sciences and Systems (CISS)}, Princeton, New
  Jersey, USA, March 2006, pp. 259--264.

\bibitem{mackay99}
D.~MacKay, ``Good error-correcting codes based on very sparse matrices,''
  \emph{IEEE Transactions on Information Theory}, vol.~45, no.~2, pp. 399--431,
  March 1999.

\bibitem{richardsonetal01a}
T.~Richardson, M.~Shokrollahi, and R.~Urbanke, ``Design of capacity-approaching
  irregular low-density parity-check codes,'' \emph{IEEE Transactions on
  Information Theory}, vol.~47, no.~2, pp. 619--637, February 2001.

\bibitem{richardsonetal01}
T.~Richardson and R.~Urbanke, ``The capacity of {LDPC} codes under message
  passing decoding,'' \emph{IEEE Transactions on Information Theory}, vol.~47,
  no.~2, pp. 599--618, February 2001.

\bibitem{rosenthaletal01}
J.~Rosenthal and P.~Vontobel, ``Constructions of regular and irregular {LDPC}
  codes using {R}amanujan graphs and ideas from {M}argulis,'' in
  \emph{Proceedings of the IEEE International Symposium on Information Theory
  (ISIT)}, Washington, District of Columbia, USA, June 24--29 2001, p.~4.

\bibitem{mcgregoretal07}
A.~McGregor and O.~Milenkovic, ``On the hardness of approximating stopping and
  trapping sets in {LDPC} codes,'' in \emph{Proceedings of the IEEE Information
  Theory Workshop (ITW)}, Lake Tahoe, California, USA, September 2007.

\bibitem{milenkovicetal07}
O.~Milenkovic, E.~Soljanin, and P.~Whiting, ``Asymptotic spectra of trapping
  sets in regular and irregular {LDPC} code ensembles,'' \emph{IEEE
  Transactions on Information Theory}, vol.~53, no.~1, pp. 39--55, January
  2007.

\bibitem{laendneretal05}
S.~Laendner and O.~Milenkovic, ``Algorithmic and combinatorial analysis of
  trapping sets in structured {LDPC} codes,'' in \emph{Proceedings of the
  International Conference on Wireless Networks, Communications, and Mobile
  Computing (WirelessComm)}, Maui, Hawaii, June 2005.

\bibitem{hollmannetal07}
H.~Hollmann and L.~Tolhuizen, ``On parity-check collections for iterative
  erasure decoding that correct all correctable erasure patterns of a given
  size,'' \emph{IEEE Transactions on Information Theory}, vol.~53, no.~2, pp.
  823--828, February 2007.

\bibitem{hehnetal06}
T.~Hehn, S.~Laendner, O.~Milenkovic, and J.~B. Huber, ``The stopping redundancy
  hierarchy of cyclic codes,'' in \emph{Proceedings of the 44th Annual Allerton
  Conference on Communication, Control and Computing}, Monticello, Illinois,
  USA, September 2006, pp. 1271--1280.

\bibitem{laendneretal06}
S.~Laendner, T.~Hehn, O.~Milenkovic, and J.~Huber, ``When does one redundant
  parity-check equation matter?'' in \emph{Proceedings of the 49th annual IEEE
  Global Telecommunications Conference (GlobeCom)}, San Francisco, California,
  USA, November 2006.

\bibitem{alonetal00}
N.~Alon and J.~Spencer, \emph{The Probabilistic Method}, ser. Interscience
  Series in Discrete Mathematics and Optimization.\hskip 1em plus 0.5em minus
  0.4em\relax John Wiley, 2000.

\bibitem{hofri95}
M.~Hofri, \emph{Analysis of Algorithms}.\hskip 1em plus 0.5em minus 0.4em\relax
  Oxford University Press, 1995.

\bibitem{brockwell64}
P.~Brockwell, ``An asymptotic expansion for the tail of the binomial
  distribution and its application in queuing theory,'' \emph{Journal of
  Applied Probability}, vol.~1, no.~1, pp. 163--169, June 1964.

\bibitem{laendneretal07}
S.~Laendner and O.~Milenkovic, ``Codes based on latin squares: Cycle structure,
  stopping set, and trapping set analysis,'' \emph{IEEE Transaction on
  Communications}, vol.~55, no.~2, pp. 303--312, February 2007.

\bibitem{hu}
\BIBentryALTinterwordspacing
X.-Y. Hu. Source code for approximating the {M}in{D}ist problem of {LDPC}
  codes. Error-Correcting Codes Website by D. MacKay. [Online]. Available:
  \url{http://www.inference.phy.cam.ac.uk/mackay/MINDIST_ECC.html}
\BIBentrySTDinterwordspacing

\bibitem{huetal04}
X.-Y. Hu, M.~Fossorier, and E.~Eleftheriou, ``On the computation of the minimum
  distance of low-density parity-check codes,'' in \emph{Proceedings of the
  IEEE International Conference on Communications (ICC)}, Paris, France, June
  2004.

\bibitem{margulis82}
G.~Margulis, ``Explicit constructions of graphs without short cycles and low
  density codes,'' \emph{Combinatorica}, vol.~2, no.~1, pp. 71--78, March 1982.

\bibitem{macwilliamsetal77}
F.~MacWilliams and N.~Sloane, \emph{The Theory of Error-Correcting
  Codes}.\hskip 1em plus 0.5em minus 0.4em\relax North-Holland Publishing
  Company, 1977.

\bibitem{kouetal01}
Y.~Kou, S.~Lin, and M.~Fossorier, ``Low-density parity-check codes based on
  finite geometries: a rediscovery and new results,'' \emph{IEEE Transactions
  on Information Theory}, vol.~47, no.~7, pp. 2711--2736, November 2001.

\bibitem{hirschfeld79}
J.~Hirschfeld, \emph{Projective geometeries over finite fields}.\hskip 1em plus
  0.5em minus 0.4em\relax Oxford Mathematical Monographs, 1979.

\bibitem{hirschfeld85}
------, \emph{Finite projective spaces of three dimensions}.\hskip 1em plus
  0.5em minus 0.4em\relax Oxford Mathematical Monographs, 1985.

\end{thebibliography}
\end{document}